\documentclass[11pt]{article}
\usepackage{amsmath,epsfig,wrapfig}
\begin{document}
\pagestyle{empty}

\begin{flushleft}
\Large
{SAGA-HE-135-98
\hfill May 27, 1998}  \\
\end{flushleft}
 
\vspace{2.0cm}
 
\begin{center}
 
\LARGE{{\bf Studies of structure functions}} \\
\vspace{0.2cm}

\LARGE{{\bf at a low-energy facility  }} \\

\vspace{1.1cm}
 
\LARGE
{S. Hino and S. Kumano $^*$ }         \\
 
\vspace{0.4cm}
  
\LARGE
{Department of Physics}         \\
 
\vspace{0.1cm}
 
\LARGE
{Saga University}      \\
 
\vspace{0.1cm}

\LARGE
{Saga 840-8502, Japan} \\

\vspace{1.4cm}
 
\LARGE
{Talk given at the Workshop on} \\

\vspace{0.2cm}

{``Future Plan at RCNP"} \\

\vspace{0.4cm}

{Osaka, Japan, March 9 -- 10, 1998} \\

{(talk on Mar. 9, 1998) }  \\
 
\end{center}
 
\vspace{1.3cm}

\vfill
 
\noindent
{\rule{6.0cm}{0.1mm}} \\
 
\vspace{-0.3cm}
\normalsize
\noindent
{* Email: 97sm16@edu.cc.saga-u.ac.jp, kumanos@cc.saga-u.ac.jp.} \\

\vspace{-0.6cm}
\noindent
{\ \ \ Information on their research is available 
 at http://www.cc.saga-u.ac.jp/saga-u}  \\

\vspace{-0.6cm}
\noindent
{\ \ \ /riko/physics/quantum1/structure.html.} \\

\vspace{+0.5cm}
\hfill
{\large to be published in proceedings}

\vfill\eject
\setcounter{page}{1}
\pagestyle{plain}
\begin{center}
 
\Large
{\bf Studies of structure functions at a low-energy facility} \\
 
\vspace{0.5cm}
 
{S. Hino and S. Kumano $^*$}             \\
 
{Department of Physics, Saga University}      \\

{Honjo-1, Saga 840-8502, Japan} \\

\vspace{0.7cm}

\normalsize
Abstract
\end{center}
\vspace{-0.60cm}

\begin{center}
\begin{minipage}[t]{10.0cm}
We discuss the studies on structure functions at the possible
future RCNP facility. At this stage, an electron-proton or 
proton-proton collider with $\sqrt{s}=5 \sim 10$ GeV is considered.
We explain large-$x$ physics, nuclear modification of sea-quark and
gluon distributions, and tensor spin structure function as the interesting
topics at the facility. The large-$x$ parton distributions
are important for finding new physics beyond QCD in anomalous events 
such as the CDF jet data.
The nuclear parton distributions are valuable in detecting
a quark-gluon signature in heavy-ion reactions.
The tensor structure function $b_1$ is a new field
of high-energy spin physics. Considering these physics possibilities,
we believe that the possible RCNP facility is important for
the hadron-structure community.
\end{minipage}
\end{center}

\vspace{0.1cm}
\section{Introduction}\label{intro}
\vspace{-0.1cm}

The purpose of this talk is to discuss interesting topics
which could be investigated by the possible future 
RCNP (Research Center for Nuclear Physics) facility
in the field of structure functions. At this stage, we consider
a collider with 10 GeV electron and a few GeV proton or the one
with a few GeV proton and a few GeV proton (or nucleus), so that
the center-of-mass energy is $\sqrt{s}=5 \sim 10$ GeV.
However, the energy range could vary depending on the physics interest.

Structure functions in the nucleon have been investigated
since the 1960s through various high-energy lepton and hadron 
scattering processes. Now the unpolarized parton distributions are
relatively well known from very small $x$ ($\sim 10^{-5}$) to large $x$
except for the gluon distribution at very small $x$ and at large $x$.
As primary future projects, high-energy facilities are discussed for
measuring the polarized structure functions such as the 
RHIC (Relativistic Heavy Ion Collider) 
and the polarized HERA (Hadron-Electron Ring Accelerator). 
Because the energy at the future RCNP facility
is expected to be much smaller than these US and European facilities,
we should focus on the large-$x$ part.

The Bjorken $x$ is related to the square of the momentum
transfer $q$ ($q^2=-Q^2$) as $x=Q^2/(2p\cdot q)$,
where $p$ is the proton momentum. Using the variable
$y=p\cdot q/(p\cdot k)$ with the initial electron momentum $k$, we rewrite
the relation as $x=Q^2/(2yp\cdot k)\approx Q^2/(ys)$ with $s=(p+k)^2$.
In order to be deep inelastic scattering, $Q^2$ has to be large
enough: typically $Q^2>1$ GeV$^2$. If the c.m. energy is
$\sqrt{s}=10$ GeV, the minimum $x$ is then given by
$x_{min}\sim 1/(10)^2=0.01$. 
The $x$ region ($0.01<x<1$) is considered as a ``large"-$x$ one
in comparison with the HERA kinematical range ($x_{min}\sim 10^{-5}$).
Therefore, we should find interesting topics in this large-$x$ region.
Even though this region has been investigated for a long time,
there are still important issues. In particular,
if accurate experimental data are taken, the large-$x$ could
be more important than the small-$x$ part which has been paid attention
to in the last several years. The large-$x$ parton
distributions have not been measured accurately at the existing facilities;
however, they are essential, for example, in explaining the CDF anomalous
jet events. In this sense, a rather low-energy but high-intensity
accelerator is crucial for finding new physics beyond quantum
chromodynamics (QCD). We discuss the importance of large-$x$ physics
in section \ref{large-x}.

The European Muon Collaboration (EMC) experimental results in 1983
shed light on nuclear modification of the parton distributions.
The modification mechanisms of the structure function $F_2$ were
studied in the medium-$x$ region for explaining the EMC results.
Then, the small-$x$ region was investigated as nuclear shadowing.
Now, the details of the $F_2$ modification are known from small $x$
to large $x$. However, nuclear sea-quark and gluon distributions
are not well determined even though they are important for applications
to high-energy heavy-ion physics. We discuss interesting nuclear
parton distributions and whether they could be measured at the
low-energy facility in section \ref{nucleus}.

The last topic is on spin-dependent structure functions.
Those for the spin-1/2 proton have been studied particularly 
in the last ten years. Now, the $g_1$ structure functions have
been measured by several experimental groups, and
we have rough idea on the longitudinally polarized parton distributions.
The missing parts in the proton are the transversity structure function
$h_1$ and higher-twist ones. Because there are other future projects
to study these spin-1/2 structure functions, we had better consider
another direction in the field of spin physics.
One of the possible ideas is to investigate new structure functions 
for spin-one particles. It is known that there is
a new twist-two structure function $b_1$, which does not exist
for spin-1/2 particles, for example in the electron-deuteron
scattering. Although the HERMES collaboration will report on $b_1$
in the near future, we do not think that the results are accurate
enough to find the small quantity. The Electron Laboratory For Europe (ELFE)
is a suitable facility; however, the project is not materialized yet.
Considering these situations, the future RCNP facility could be 
the first one to measure $b_1$ if it is approved
in the near future. We discuss this point in section \ref{spin}.

The above topics are important for finding new physics at very high-energy
accelerators, for detecting a quark-gluon plasma signature, and
for creating a new field of high-energy spin physics.
Therefore, the future RCNP facility should be valuable 
in the hadron-physics community.
In the following sections, we discuss the details of each topic.

\vspace{0.1cm}
\section{Large-x physics for finding a signature beyond QCD}
\label{large-x}
\vspace{-0.1cm}

We may think that the medium and large $x$ physics has been already
investigated extensively and that no interesting physics
is left. It may be right in the sense that a lot of experimental data
exist; however, if a high-intensity facility is built, the situation could
be different. In order to convince that the large $x$ is important,
we discuss well-known CDF (Collider Detector at Fermilab)
anomalous jet events as an interesting example and their relation
to the large-$x$ gluon distribution. 

The CDF collaboration measured the inclusive jet cross sections
in the $p+\bar p$ reaction with $\sqrt{s}=1.8$ TeV.
The measured cross sections agree, in general, with
the next-to-leading-order (NLO) QCD calculation.
However, they found significant differences from the NLO prediction
in the large jet transverse-energy region, $E_T>300$ GeV \cite{cdf}.
In Fig. 1 of Ref. \cite{cdf}, they show the fractional difference
from the NLO calculation with the MRS-D0$'$ input distribution.
They also show theoretical predications with different parton
distributions: MRSA$'$, MRSG, CTEQ2M, CTEQ2ML, and GRV-94. 
Even though the theoretical cross sections
depend on the distribution model, the variations are within
about 10\%. On the other hand, the CDF data deviate about 30\%
in the large $E_T$ region. It was thought to be much larger than
theoretical ambiguities. 

\begin{wrapfigure}{r}{0.46\textwidth}
   \vspace{-0.2cm}
   \begin{center}
       \epsfig{file=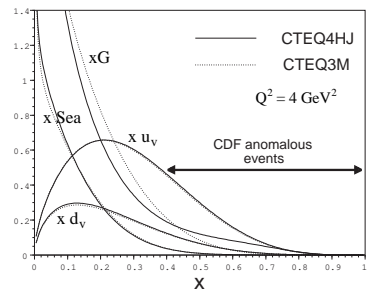,width=6.0cm}
   \end{center}
   \vspace{-0.7cm}
       \caption{\footnotesize
          The CTEQ3M and CTEQ4HJ parton distributions at $Q^2$=4 GeV$^2$.}
       \label{fig:cteq}
\end{wrapfigure}
Because this is the unexplored kinematical
region, people speculated exotic mechanisms such as subquark.
However, it became clear later according to the CTEQ collaboration
\cite{cteq4} that the anomalous jet events could be
explained if the gluon distribution is significantly larger
than the CTEQ2M and CTEQ3M distributions at $x>0.4$. 
In Fig. \ref{fig:cteq}, the CTEQ3M and CTEQ4HJ parton distributions
are shown by the dotted and solid curves, respectively.
It is obvious from the figure that the quark and antiquark
distributions are essentially the same in the two parametrizations.
However, the gluon distributions differ significantly.
The CDF events are taken at large $E_T$ so that the distributions
should be evolved to the scale $Q^2=(E_T/2)^2$. In the central
rapidity region, the contributing partons have the fraction of momentum,
$x_{1,2}\sim 2 E_T/\sqrt{s}$. Substituting $\sqrt{s}=1.8$ TeV
and for example $E_T \sim 350$ GeV, we obtain $x_{1,2}\sim 0.4$.
The major subprocesses at such a large $E_T$ are
quark-quark and quark-gluon interactions, so that the parton
distributions should be supplied at large $x$ and large $Q^2$.
The standard way is to use the parton distributions, which are optimized
so as to explain many other experimental data, then to evolve them to
large $Q^2$ by using the DGLAP equations:
\begin{align}
\! \! \! \! \! \! \! \! \! \! \! \! \! \! \! \!
\frac{\partial}{\partial (ln \, Q^2)} \, q(x,Q^2) & = 
\frac{\alpha_s}{2\pi}
\int_x^1 \frac{dy}{y}\, \left[ \, P_{qq} (x/y)\, q(y,Q^2)
                              +\  P_{qG} (x/y)\, G(y,Q^2)\, 
                       \right]
\, ,
\nonumber \\
\frac{\partial}{\partial (ln \, Q^2)} \, G(x,Q^2) & = 
\frac{\alpha_s}{2\pi}
\int_x^1 \frac{dy}{y}\, \left[ \, P_{Gq} (x/y)\, q(y,Q^2)
                              +   P_{GG} (x/y)\, G(y,Q^2)\,
                       \right]
\, .
\label{eqn:dglap}
\end{align}
According to these evolution equations, the parton distributions
in the $x$ region, $0.4\le x \le 1$, should be known at a certain
low $Q^2$ in order to calculate those distributions at very large
$Q^2$ ($= E_T^2/4$). However, the gluon distribution in this $x$
region is not known at all as obvious from Fig. \ref{fig:cteq},
where the CTEQ4HJ gluon distribution is much larger than the
CTEQ3M one. It should be noted that the CTEQ4HJ is in the perfect
agreement with the CDF data on the contrary to the CTEQ3M. 

\begin{wrapfigure}{r}{0.46\textwidth}
   \vspace{-0.2cm}
   \begin{center}
       \epsfig{file=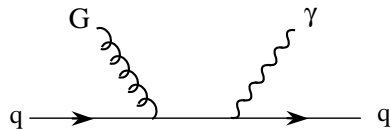,width=5.5cm}
   \end{center}
   \vspace{-0.4cm}
       \caption{\footnotesize
          Direct photon process for finding 
                        the large-$x$ gluon distribution.}
   \vspace{0.4cm}
       \label{fig:gamma}
\end{wrapfigure}
In this way, we find that the large-$x$ gluon distribution
is essential for determining whether or not the CDF events are
really anomalous. At this stage, there is no way to fix the
gluon distribution at such a large $x$. In order to confirm
the conservative CTEQ explanation, we have to measure the
gluon distribution. Although the high-energy accelerator is suitable
for studying the small $x$ distributions,
the low-energy facility like the future RCNP is
valuable for the large-$x$ measurements. We should be, however,
careful that the intensity is high enough to find
small quantities. The large-$x$ gluon distribution could
be probed by the direct photon process in Fig. \ref{fig:gamma}, 
but the details should be studied on higher-order corrections (K-factor),
higher-twist effects, possible photon background, and expected
hard-photon $p_{_T}$ distribution. Because the RCNP energy is not
fixed yet, we may study the optimum one for measuring the gluon
distribution in the direct photon process. The c.m. energy
$\sqrt{s}\sim 10$ GeV may not be large enough for the direct
photon process.

\vfill\eject
\section{Parton distributions in nuclei}\label{nucleus}
\vspace{-0.1cm}

\begin{wrapfigure}{r}{0.46\textwidth}
   \vspace{0.0cm}
   \begin{center}
       \epsfig{file=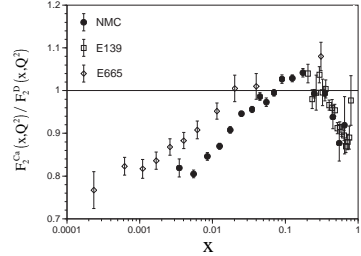,width=5.5cm}
   \end{center}
   \vspace{-0.8cm}
       \caption{\footnotesize
          Experimental data for $F_2^{Ca}/F_2^D$.}
   \vspace{0.4cm}
       \label{fig:f2a}
\end{wrapfigure}
The parton distributions are modified in the nuclear environment,
and the modification is well investigated through the structure
function $F_2$. As an example, the ratio $F_2^{Ca}/F_2^D$ is shown
in Fig. \ref{fig:f2a}, where the SLAC-E139, New Muon Collaboration (NMC),
and Fermilab-E665 data are included \cite{f2a}. 
The $F_2$ structure function has been measured also for various size nuclei. 
The large-$x$ region ($x>0.7$) is usually attributed to the nucleon
Fermi motion in the nucleus, the medium-$x$ to the binding mechanism
and confinement-radius change, and the small-$x$ to the nuclear shadowing.
It is not the purpose of this paper to discuss the details of these
mechanisms, so that the interested reader may read a summary paper
\cite{summary} or a recent report \cite{ku}.

Although there are some differences between the NMC and E665 data,
the structure functions $F_2^A$ have been rather well studied
from very small $x$ to large $x$. The RCNP energy range is, at least
at this stage, close to the one for the fixed target experiments at SLAC.
Because the SLAC group has done the extensive studies of nuclear $F_2$,
we had better think about other possibilities.
To know the $F_2$ structure function in a nucleus does not mean
that all the parton distributions are known in the nucleus.
The $F_2$ structure function is given by
$F_2 = x \sum_i e_i^2 (q_i +\bar q_i)$.
It is dominated by the sea-quark distributions at small $x$ ($x<0.01$)
and by the valence-quark ones at large $x$ ($x>0.3$). This fact means
that the sea (valence) quark modification at small (large) $x$ is known
from the $F_2$ measurements. However, the modification of the valence
and sea quark distributions is not known in the whole-$x$ range.
The sea-quark distributions in the proton are determined
by using various experimental data such as electron/muon
deep inelastic scattering, neutrino scattering, Drell-Yan process,
and $W$ production cross sections. In the nuclear case, a variety
of these experimental data are not available at this stage, so that
the precise determination of each quark/antiquark distribution
is not possible.

\begin{figure}[h]
\parbox[t]{0.46\textwidth}{
   \begin{center}
       \epsfig{file=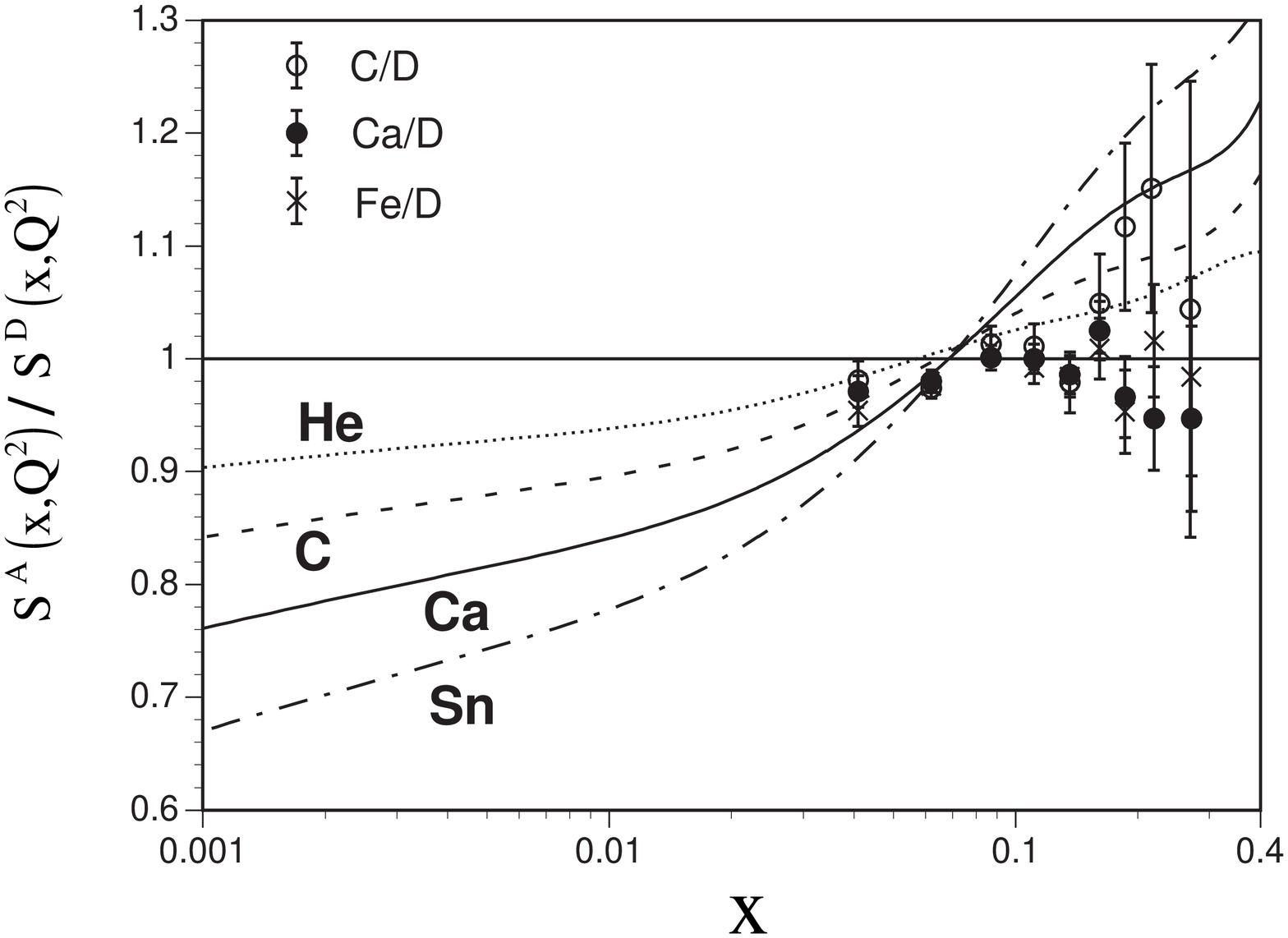,width=6.0cm}
   \end{center}
   \vspace{-0.8cm}
       \caption{\footnotesize
          Sea-quark distributions in nuclei.}
       \label{fig:sea}
}\hfill
\parbox[t]{0.46\textwidth}{
   \begin{center}
   \epsfig{file=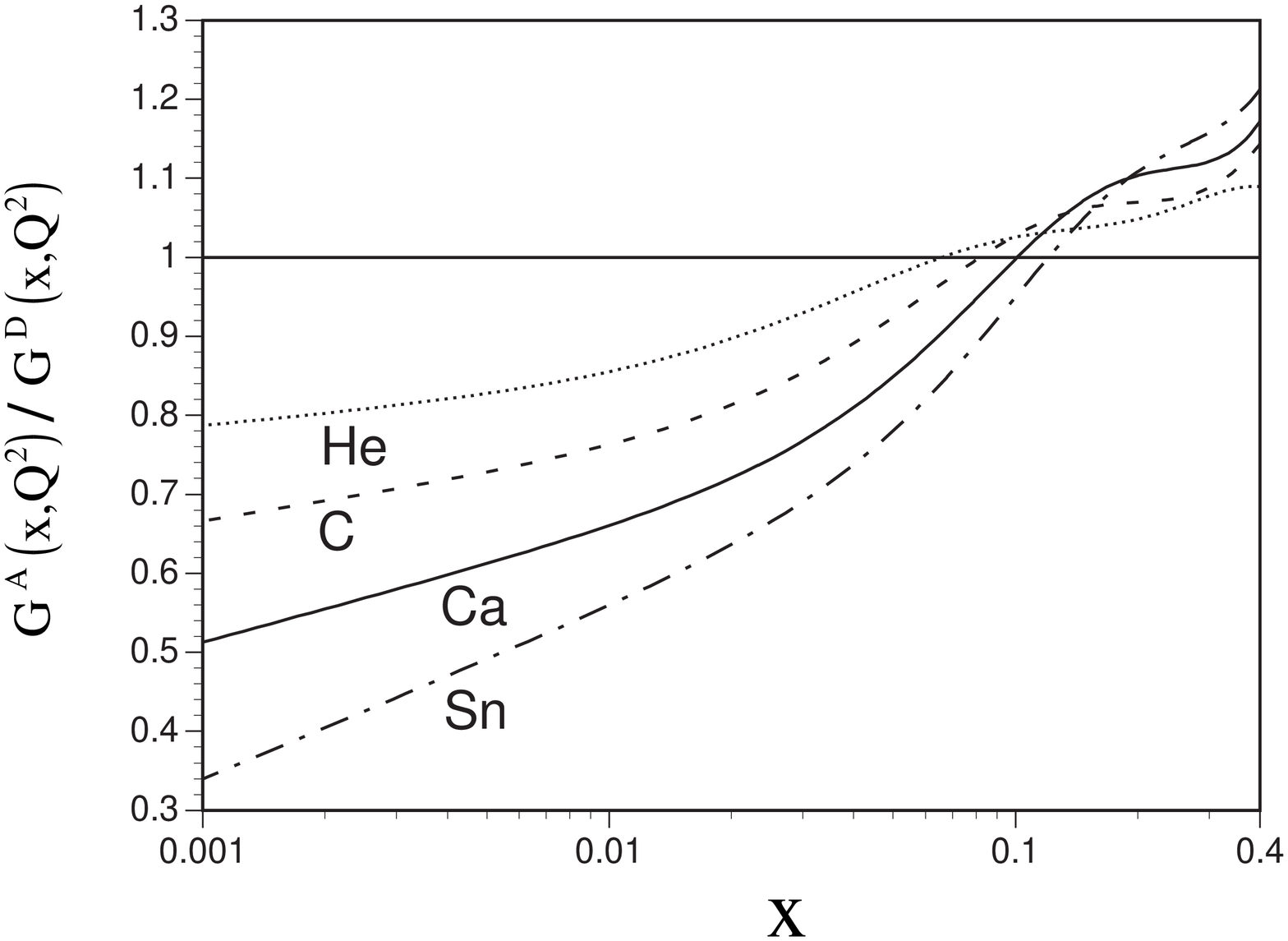,width=6.0cm}
   \end{center}
   \vspace{-0.8cm}
       \caption{\footnotesize 
          Gluon distributions in nuclei.}
       \label{fig:glue}
}
\end{figure}

Considering the above situation, we think that the interesting
future direction is to separate valence and sea quark distributions.
Then, each flavor distribution should be also determined.
Furthermore, it is noteworthy that little is known for the gluon
distributions in nuclei although they play a major role in high-energy
heavy-ion reactions.
As an example of model predictions, we show the sea-quark and gluon
distributions for the nuclei He, C, Ca, and Sn in Figs.
\ref{fig:sea} and \ref{fig:glue}. The distributions are calculated
at $Q^2=5$ GeV$^2$ in a parton model with $Q^2$ rescaling
and parton-recombination mechanisms \cite{ku}.
The model parameters are determined so that the theoretical ratio
agrees with the $F_2^{Ca}/F_2^D$ data.
For the details of the model, the reader is suggested to read
Ref. \cite{ku}. Here, the rescaling model is employed as an effective
model which includes the binding-type nuclear effects and
confinement-size modification, and the recombination model
is as a shadowing model in an infinite momentum frame.

\vspace{-0.8cm}
\noindent
\vspace{0.1cm}

Because the future RCNP cannot compete with other high-energy facilities
in the small-$x$ physics, we should think about possible physics in the
$x$ region, $x>0.05$.
The sea-quark shadowing becomes conspicuous in the $x$ region,
$x<0.01$, so that higher energy facility should be appropriate
for measuring the sea shadowing. In fact, it will be investigated at RHIC.
The E772 Drell-Yan data \cite{e772} are also shown in Fig. \ref{fig:sea}.
Although the iron data are often quoted in suggesting that
there is no sea-quark modification in the $x\sim 0.1$ region,
the situation is not so clear: the carbon data lie above the unity
and the calcium data are below. It is hard to believe this kind
of A dependence. Because the sea-quark distribution itself is
very small at $x>0.2$, the experimental errors become large.
Fortunately, this region is just the kinematical range of the future RCNP
facility. If it has enough intensity, it should be possible
to measure the sea modification in detail, particularly the A dependence.
Because the sea-quark distribution in this $x$ range cannot be determined
by $F_2$, we should rely on the Drell-Yan process 
$p+A\rightarrow \mu^+ \mu^- +X$ for various nuclei. 
The accurate measurements should clarify theoretical issues. 
For example, although it is not explicitly included in the above model, 
the sea-quark enhancement is generally predicted in the pion-excess model
in contradiction to the E772 iron data.
Therefore, future RCNP data could shed light on the pion-excess mechanism,
namely on the nuclear force. 
The accurate data will be valuable also 
for determining the nuclear parton distributions in the whole-$x$ range. 

The prediction for the nuclear gluon distribution is shown in
Fig. \ref{fig:glue}. Although there are implicit data on
the gluon modification, there is no accurate explicit data
at this stage. The gluon shadowing takes place at $x\sim 0.1$
and it becomes conspicuous at $x\sim 0.01$.
Since the gluon shadowing will be investigated also at RHIC, 
the future RCNP may address the region $x>0.1$. In the nucleon case,
the scaling violation of $F_2$, direct-photon process,
and J/$\psi$ production are used for determining its gluon distribution.
The wide range of scaling violation data and the significant
J/$\psi$ events would not be obtained at the RCNP, 
so that the remaining possibility is to use the direct photon process.
It has been already discussed in section \ref{large-x}.
Because there may exist complexities due to the available low energy,
we should study the reaction in detail.

If the nuclear sea-quark and gluon distributions are obtained
at the future RCNP in the $x$ range, $x>0.1$, they should be valuable
not only for establishing the theoretical nuclear model but also for
applications to high-energy heavy-ion physics. In particular,
we believe that the accurate nuclear parton distributions are essential
for finding a quark-gluon plasma signature.
For example, although the J/$\psi$ production may be related to
such a signature, its cross section is not precisely calculated
at this stage due to the lack of information on the nuclear gluon
distributions. 

\vspace{0.1cm}
\section{Polarized parton distributions}\label{spin}
\vspace{-0.1cm}

Spin-dependent structure functions are studied for the proton and
for the ``neutron". Now, there are many data on the structure function $g_1$.
However, as the $F_2$ measurements could not fix the valence and sea quark
distributions in section \ref{nucleus}, each polarized distribution cannot
be determined solely by the $g_1$ data. Therefore, we have to wait
for future measurements to solve the proton spin issue completely
\cite{saga}. As it was explained in section \ref{large-x},
the large-$x$ unpolarized distributions could be studied at the future
RCNP. In the same way, the polarized distributions at large $x$ should be
important for finding an exotic signature in polarized reactions.
It could be one of the interesting topics on spin physics at the facility.
However, there are future plans and proposals at BNL, CERN, DESY, and SLAC
on the polarized parton distributions for the spin-1/2 nucleon,
so that we had better focus on a different spin topic. 
One of the other possibilities is to investigate
a new spin-dependent structure function $b_1$ in the spin-one hadrons. 

The polarized deuteron is used for measuring the $g_1$ structure
function of the neutron. Because the deuteron is a spin-one hadron,
there should exist extra spin-dependent structure functions.
These are named $b_1$, $b_2$, $b_3$, and $b_4$ \cite{b1}.
Because the twist-two ones are related by the Callan-Gross type
relation $b_2=2xb_1$ in the leading order, the essential part is
to study $b_1$ or $b_2$. It is also interesting to investigate
a quadrupole sum rule for $b_1$ \cite{ck}. 
The $b_3$ and $b_4$ are higher-twist structure
functions, so that it is not worth while discussing the details at this stage.
These structure functions are defined in the hadron tensor $W_{\mu\nu}$
in the polarized electron-deuteron reaction; however, the expression
is too lengthy to write it down here.  The reader may look at Ref.
\cite{b1} or \cite{elfeb1}.

The $b_1$ structure function is discussed within the context
of the ELFE proposal in Ref. \cite{elfeb1}. If the RCNP facility
is intense enough, it could be used for measuring $b_1$
which is expected to be very small.
In order to measure $b_1$, the electron does not have to be
polarized. It is related to the polarized cross sections by
\begin{equation}
b_1 \propto d\sigma(0) - \frac{d\sigma(+1) + d\sigma(-1)}{2}
\ ,
\label{eqn:b1}
\end{equation}
where $d\sigma(H)$ indicates the electron-deuteron cross section
with the $z$-component $H$ of the target spin.
Combining the cross sections with a target polarized
parallel (and antiparallel) to the lepton beam direction
with the unpolarized cross section, we obtain $b_1$. 
It can be expressed also in the parton model.
Calculating the cross sections in a parton model, we have
the expression 
\begin{align}
b_1 (x) \, = \,  &
             \sum_i e_i^2 \, 
            [ \, \delta q_i(x)+ \delta \bar q_i(x) \, ] 
\ , 
\nonumber \\
\ & \delta q_i (x) \, = \,  {q_\uparrow ^0} _i (x) 
          - \frac{1}{2}
     [ {q_{\uparrow i} ^{+1} } (x) + {q_{\uparrow i} ^{-1}} (x) ]   
           \, = \,   \frac{1}{2} [{q_i ^0}  (x) 
                            -{q_{i} ^{+1} } (x)]    
\ ,
\label{eqn:b1-parton}
\end{align}
where the superscript indicates the hadron helicity
in an infinite momentum frame. 

As it is obvious from Eqs. (\ref{eqn:b1}) and
(\ref{eqn:b1-parton}),  $b_1$ is related to the tensor structure
of the deuteron. It is well known that the D-state admixture
gives rise to the finite quadrupole moment of the deuteron.
The $b_1$ structure is related to such physics. Of course,
the deep inelastic process is under consideration right now,
so that the electric quadrupole structure probed by $b_1$ could
be very different from ordinary low-energy results.
In this sense, $b_1$ is a suitable structure function
which could indicate ``exotic components" of the hadron structure.

There is an interesting point on the sum rule.
The Gottfried sum rule has been studied well last several years,
and its failure resulted in revealing the light antiquark flavor asymmetry.
A similar sum rule exists for $b_1$ according to Ref. \cite{ck}.
The similarity is obvious if they are written together:
\begin{alignat}{3}
& \! \! \! \! \! \! \! \! \! \! \text{Gottfried:} & \ & \ & \ & \  
\nonumber \\
                        & \int dx \, [F_2^p(x)-F_2^n(x)] & \, = \, &
                        \frac{1}{3} &  & \, + \, \frac{2}{3} 
                        \int dx \, [ \, \bar u(x)-\bar d(x) \, ] 
\label{eqn:sumgot}
\ , \\
& \! \! \! \! \! \! \! \! \! \! \text{Ref.} \, \text{\cite{ck}:}
                        \ \ \ \ \ \ \ \ \ \,  
                         \int dx \, b_1 (x) & \, = \, & 
                        {\displaystyle \lim_{t\rightarrow 0}}
                        - \frac{5}{3} \frac{t}{4M^2} F_Q (t) &
                        & \, + \, \frac{1}{9} \, \delta Q_{sea}  
\ ,
\label{eqn:sumb1}
\end{alignat}
where $\delta Q_{sea}$ is the sea-quark tensor polarization,
for example
$\delta Q_{sea}= \int dx  
          [8 \delta \bar u (x) +2 \delta \bar d (x)
           +\delta s (x) +\delta \bar s (x)]$ 
for the deuteron, and
$F_Q (t=0)$ is the quadrupole moment in the unit
of $e/M^2$ for a spin-one hadron with the mass $M$.
As it is shown in Eqs. (\ref{eqn:sumgot}) and (\ref{eqn:sumb1}), 
there are following similarities.
Because the valence-quark number depends on flavor,
the finite sum 1/3 is obtained in the Gottfried sum rule.
However, the first term vanishes in the $b_1$ case, which 
reflects the fact that the valence number does not depend on spin.
The second term in Eq. (\ref{eqn:sumb1}) corresponds 
to $\int dx (\bar u-\bar d)$ in Eq. (\ref{eqn:sumgot}).
If a deviation from the sum $\int dx b_1(x)=0$ is found,
it should suggest a finite sea-quark tensor polarization
as the Gottfried-sum-rule violation suggested
a finite $\bar u-\bar d$ distribution.

The theoretical study of $b_1$ is still at the preliminary stage,
and there exists no experimental data. The HERMES collaboration
will report on $b_1$ in a few years. However, 
because the $b_1$ is expected to be very small, $b_1/F_1 \sim 0.01$
in a naive quark model for the deuteron \cite{elfeb1}, they would not
be able to measure it. It may be possible at ELFE, but the facility
itself is not approved yet. If the RCNP facility will be built in the
near future, it could be the first one to measure the tensor structure
function $b_1$. This is a new field of high-energy spin physics so
that unexpected experimental results could be obtained.
The studies on the new spin structure are important
for testing our knowledge of high-energy spin physics
in the unexplored field and for establishing the hadron structure model
in the high-energy region.

\vspace{0.1cm}
\section{Summary}\label{summary}
\vspace{-0.1cm}

We have discussed the studies of structure functions at the possible
future RCNP facility, which is considered as a ``low-energy"
facility in comparison with those at BNL, CERN, DESY, and Fermilab.
First, we focused on the large-$x$ part of parton distributions.
Explaining the CDF anomalous events and their relation to
the large-$x$ parton distributions, we concluded that
the large-$x$ physics is important for finding new
physics beyond quantum chromodynamics. A rather low-energy
machine with high intensity is suitable for measuring the
parton distributions at large $x$.
Second, possible studies on nuclear parton distributions
were discussed. In particular, modification of sea-quark
and gluon distributions is not well known although it is
important for finding the quark-gluon plasma signature.
Third, spin-dependent structure functions were discussed.
The large-$x$ part could be also studied in the same way
as the unpolarized case; however, the tensor structure
function $b_1$ should be an interesting one as a new
topic in high-energy spin physics.
From these discussions, we think that it is worth while
proposing the new RCNP facility.

 
\vspace{0.4cm}

\noindent
{* Email: 97sm16@edu.cc.saga-u.ac.jp, kumanos@cc.saga-u.ac.jp.} \\

\vspace{-0.55cm}
\noindent
{\ \ \ Information on their research is available 
 at http://www.cc.saga-u.ac.jp/saga-u}  \\

\vspace{-0.55cm}
\noindent
{\ \ \ /riko/physics/quantum1/structure.html.} 

\vspace{0.1cm}


\begin{thebibliography}{99}
\bibitem{cdf}     F. Abe et al. (CDF collaboration), 
                          Phys. Rev. Lett. 77 (1996) 438.
\vspace{-0.2cm}
\bibitem{cteq4}   H. L. Lai et. al. (CTEQ collaboration),
                          Phys. Rev. D55 (1997) 1280.
\vspace{-0.2cm}
\bibitem{f2a}     J. Gomez et al. (SLAC-E139 collaboration),
                          Phys. Rev. D49 (1994) 4348;
                  P. Amaudruz et al. (New Muon Collaboration), 
                          Z. Phys. C51 (1991) 387;
                          Nucl. Phys. B441 (1995) 3;
                  M. R. Adams et al. (Fermilab-E665 collaboration), 
                          Phys. Rev. Lett. 68 (1992) 3266;
                          Z. Phys. C67 (1995) 403.
\vspace{-0.2cm}
\bibitem{summary} D. F. Geesaman, K. Saito, and A. W. Thomas,
                          Ann. Rev. Nucl. Part. Sci. 45 (1995) 337.
\vspace{-0.2cm}
\bibitem{ku}      S. Kumano and K. Umekawa, 
                          SAGA-HE-130-98 (hep-ph/9803359).
\vspace{-0.2cm}
\bibitem{e772}    D. M. Alde et al. (Fermilab-E772),
                          Phys. Rev. Lett. 64 (1990) 2479.
\vspace{-0.2cm}
\bibitem{saga} S. Hino, M. Hirai, S. Kumano, and M. Miyama,
                          research in progress. 
\vspace{-0.2cm}
\bibitem{b1}     P. Hoodbhoy, R. L. Jaffe, and A. Manohar,
                          Nucl. Phys. B312 (1989) 571.
\vspace{-0.2cm}
\bibitem{ck}      F. E. Close and S. Kumano,
                          Phys. Rev. D42 (1990) 2377.
\vspace{-0.2cm}
\bibitem{elfeb1}  S. Kumano, pp. 371 in THE ELFE PROJECT
                  ``an Electron Laboratory for Europe",
                  edited by J. Arvieux and E. De Sanctis,
                  Italian Physical Society,
                  Conference Proceedings Vol. 44 (1993).
\end{thebibliography}
\end{document}